\documentclass[reprint,aps,prb,twocolumn,superscriptaddress]{revtex4-2}

\usepackage{graphicx}
\usepackage{dcolumn}
\usepackage{bm}
\usepackage{amsmath}
\usepackage{braket}
\usepackage[english]{babel}
\usepackage[utf8]{inputenc}
\usepackage[dvipsnames]{xcolor}
\usepackage{soul}
\usepackage{units}
\colorlet{RED}{red}

\usepackage{hyperref}
\allowdisplaybreaks
\begin{document}
\title{Modelling of planar germanium hole qubits in electric and magnetic fields}

\author{Chien-An Wang}
\affiliation{QuTech and Kavli Institute of Nanoscience, Delft University of Technology, PO Box 5046, 2600 GA Delft, The Netherlands}
\author{H.\ Ekmel Ercan}
\affiliation{Electrical and Computer Engineering Department, University of California, Los Angeles, California 90095, United States}
\author{Mark F.\ Gyure}
\affiliation{Electrical and Computer Engineering Department, University of California, Los Angeles, California 90095, United States}
\affiliation{Center for Quantum Science and Engineering, University of California, Los Angeles, California 90095, United States}
\author{Giordano~Scappucci}
\affiliation{QuTech and Kavli Institute of Nanoscience, Delft University of Technology, PO Box 5046, 2600 GA Delft, The Netherlands}
\author{Menno~Veldhorst}
\affiliation{QuTech and Kavli Institute of Nanoscience, Delft University of Technology, PO Box 5046, 2600 GA Delft, The Netherlands}
\author{Maximilian~Rimbach-Russ}
\affiliation{QuTech and Kavli Institute of Nanoscience, Delft University of Technology, PO Box 5046, 2600 GA Delft, The Netherlands}

\begin{abstract}
Hole-based spin qubits in strained planar germanium quantum wells have received considerable attention due to their favourable properties and remarkable experimental progress. The sizeable spin-orbit interaction in this structure allows for efficient qubit operations with electric fields. However, it also couples the qubit to electrical noise. 
In this work, we perform simulations of a heterostructure hosting these hole spin qubits. We solve the effective mass equations for a realistic heterostructure, provide a set of analytical basis wave functions, and compute the effective g-factor of the heavy-hole ground-state. Our investigations reveal a strong impact of highly excited light-hole states located outside the quantum well on the g-factor. We find that sweet spots, points of operations that are least susceptible to charge noise, for out-of-plane magnetic fields are shifted to impractically large electric fields. However, for magnetic fields close to in-plane alignment, partial sweet spots at low electric fields are recovered. Furthermore, sweet spots with respect to multiple fluctuating charge traps can be found under certain circumstances for different magnetic field alignments. This work will be helpful in understanding and improving coherence of germanium hole spin qubits.
\end{abstract}

\maketitle

Hole spins in germanium quantum dots constitute a compelling platform for quantum computation~\cite{vandersypenInterfacingSpinQubits2017,scappucciGermaniumQuantumInformation2020}. Holes in germanium benefit from the strong spin-orbit interaction (SOI), absence of valley degeneracy and large heavy-hole and light-hole splitting~\cite{terrazosTheoryHolespinQubits2021}, small in-plane effective mass~\cite{hendrickxGatecontrolledQuantumDots2018}, and the formation of ohmic contacts with metals~\cite{watzingerGermaniumHoleSpin2018, hendrickxGatecontrolledQuantumDots2018,lodariLightEffectiveHole2019a}. These properties allowed a rapid development of planar germanium spin qubits from quantum dots~\cite{hendrickxGatecontrolledQuantumDots2018}, single and two qubit manipulation~\cite{hendrickxFastTwoqubitLogic2020}, singlet-triplet qubits~\cite{jirovecDynamicsHoleSingletTriplet2022}, to a 2x2 qubit array~\cite{hendrickxFourqubitGermaniumQuantum2021} as well as high-fidelity operations~\cite{lawrieSimultaneousSinglequbitDriving2021}, and rudimentary error correction circuits~\cite{vanriggelenPhaseFlipCode2022}.

The challenge for hole spin qubits is to overcome decoherence due to charge noise coupling through the spin-orbit interaction~\cite{stanoReviewPerformanceMetrics2022,froningStrongSpinorbitInteraction2021,wangUltrafastOperationsHole2022}. Current dephasing times are ~$T_2^\star=\unit[100]{ns}\,$-$\,\unit[10]{\mu s}$, which could be extended to $T_2=\unit[1000]{\mu s}$ using dynamical decoupling~\cite{lawrieSimultaneousSinglequbitDriving2021,hendrickxSweetspotOperationGermanium2024}.  
The possibility of extended coherence times in germanium hole qubits is studied in several theoretical works for nanowire~\cite{stanoFactorElectronsGatedefined2018,boscoHoleSpinQubits2021,michalLongitudinalTransverseElectric2021,boscoHoleSpinQubits2022} and planar systems~\cite{venitucciSimpleModelElectrical2019,wangOptimalOperationPoints2021a,adelsbergerHoleSpinQubits2021,boscoSqueezedHoleSpin2021a}. The coherence time can be greatly extended by operating at optimal operation points, so-called sweet spots, where the qubit resonance frequency has a vanishing derivative with respect to electric fields. Interestingly, it is predicted that at such sweet spots the electric dipole spin resonance (EDSR) driving is also be the most efficient~\cite{mauroGeometryDephasingSweet2024}. In this work, we investigate the existence of sweet spots in detail. We model the system based on recent experiments, considering a realistic potential profile resulting from a SiGe/Ge/SiGe heterostructure~\cite{lodariLowPercolationDensity2021}. We show that many basis wave-functions are required for predicting the susceptibility of the g-factor to electric fields~\cite{wimbauerZeemanSplittingExcitonic1994,delvecchioVanishingZeemanEnergy2020,seminaInfluenceSpinorbitSplitoff2021}, shifting predictions for sweet spots in out-of-plane magnetic fields to experimentally inaccessible electric field values. However, we also show that sweet spots with respect to electric fields in arbitrary directions can exist, when the magnetic field is applied with angle $\theta\lesssim\arctan(g_\parallel/g_\perp)/3=\unit[0.2]{^\circ}$, where $g_\parallel$ ($g_\perp$) is the bare in-plane (out of plane) g-factor of the heavy-hole state.

\begin{figure}[t] 
\centering
    {\includegraphics[width=\columnwidth]{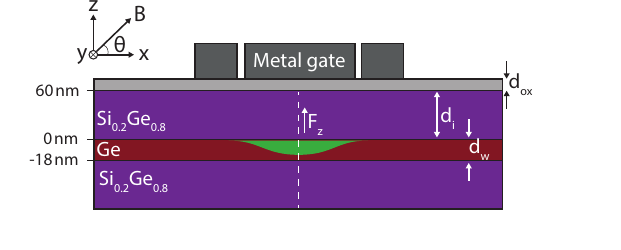}} 
    \caption{Schematics of a gate-defined quantum dot in a planar germanium heterostructure. The quantum dot is confined in the z-direction by the SiGe-Ge-SiGe layers and the Ge quantum well has width $d_w=\unit[18]{nm}$. The insulating oxide layer has width $d_{ox}=\unit[5]{nm}$. The in-plane confinement is created by the electrostatic gates which are located at the top of the heterostructure. Our model assumes a uniform electric field in the z-direction and a parabolic potential in the xy-plane. The potential profile along the dashed line is plotted in Fig.~\ref{fig:Fig2}A. The illustration of the accumulated hole wave function is colored in green. }

\label{fig:Fig1}
\end{figure}

\begin{figure*}[t] 
\centering
    {\includegraphics[width=\textwidth]{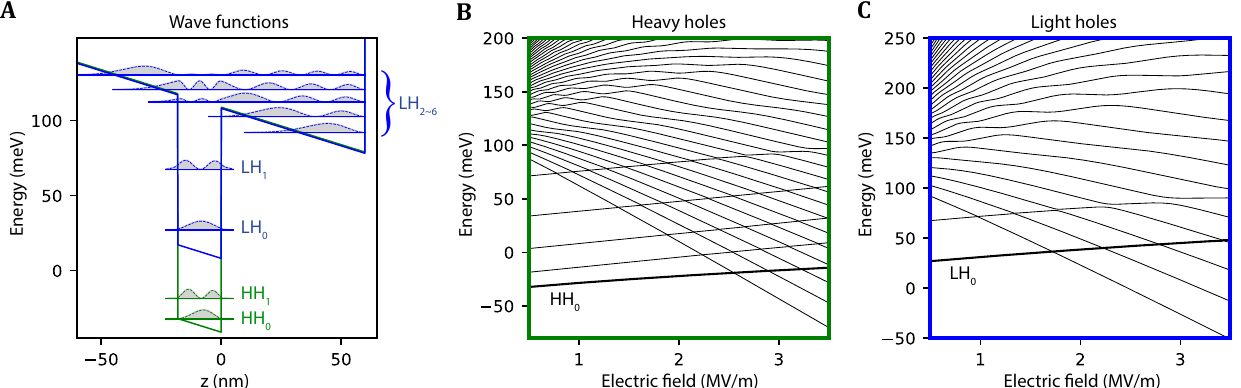}} 
    \caption{\textbf{A}, The potential of the heterostructure along the growth direction and the $\rm n^{th}$ sub-band of the heavy (light) hole levels $\rm HH_{n}$ ($\rm LH_{n}$). For this plot, the electric field strength is $F_z=\unit[0.5]{MV/m}$. \textbf{B, C}, The energy levels of the $\rm n^{th}$ heavy-hole sub-bands and the light-hole sub-bands. The levels with negative slope are located in the quantum well, while levels with a positive slope spread outside the quantum well. }

\label{fig:Fig2}
\end{figure*}

\section{Results}
In this work we describe a single hole confined vertically in a strained $\text{SiO}_{2}/\text{Si}_{0.2}\text{Ge}_{0.8}/\text{Ge}/\text{Si}_{0.2}\text{Ge}_{0.8}$ planar heterostructure using an electrostatic potential through metallic gates. Fig.~\ref{fig:Fig1} shows a sketch of the modelled device. The full Hamiltonian describing the hole reads
 \begin{align}
    H = H_\text{kin} + V_\perp(z) + V_\parallel(x,y) + H_\text{Zeeman},
    \label{eq:fullHamiltonian}
\end{align} 
where $H_\text{kin}$ is the kinetic energy operator, $V_\perp(z)$ and $V_\parallel(x,y)$ describes the vertical and planar confinement, and $H_\text{Zeeman}$ describes the interaction of the spin and the magnetic field.
\subsection{Effective mass theory for strained germanium}
Since our quantum dot structures are large compared to the inter-atom distances and operated at low densities $\rho\sim\unit[10^{10}]{cm^{-2}}$ (single hole regime), the wave-functions are localized close to the $\Gamma$ point at $\boldsymbol{k}=0$. In this regime and within the effective mass approximation, the kinetic energy is well-described by the $6\times 6$ Luttinger-Kohn Hamiltonian. Additionally, in germanium the split-off band is far separated in energy by $\Delta_\text{SO}=\unit[0.29]{eV}$ and thus negligible for the low-energy dynamics. This allows us to reduce our investigation to the standard $4\times 4$ Luttinger-Kohn Hamiltonian. In the basis of total angular momentum eigenstates $\ket{j,m_j}=\lbrace\ket{\frac{3}{2},\frac{3}{2}},\ket{\frac{3}{2},-\frac{3}{2}},\ket{\frac{3}{2},\frac{1}{2}},\ket{\frac{3}{2},-\frac{1}{2}}\rbrace$ the Luttinger-Kohn Hamiltonian reads
 \begin{align}
    H_\text{kin}=H_\text{LK} = \left(\begin{array}{cccc}
        P+Q & 0 & S & R  \\ 
        0 & P+Q & R^\dagger & -S^\dagger  \\
        S^\dagger & R & P-Q & 0  \\
        R^\dagger & -S & 0 & P-Q  
    \end{array}\right).
    \label{eq:HLK}
\end{align} 
The upper-left block $P+Q$ describe the kinetic energy of the spin-$\frac{3}{2}$ heavy-hole state, the lower-right block $P-Q$ describes the kinetic energy of the spin-$\frac{1}{2}$ light-hole state, $S$ describes the heavy-light-hole coupling with same spin, and $R$ describes the heavy-light-hole coupling with opposite spin direction. The operators are described by
 \begin{align}
    P&= \frac{\hbar^2}{2m_0}\gamma_1 (k_x^2+k_y^2+k_z^2),\\
    Q&= \frac{\hbar^2}{2m_0}\gamma_2 (k_x^2+k_y^2-2k_z^2),\\
    R&= \sqrt{3}\frac{\hbar^2}{2m_0}\left[-\gamma_2 (k_x^2-k_y^2)+i\gamma_3 k_x k_y+i\gamma_3 k_y k_x\right],\\
    S&= -\sqrt{3}\frac{\hbar^2}{2m_0}\gamma_3 \left[(k_x-i k_y)k_z+k_z(k_x-i k_y)\right],
    \label{eq:LK}
\end{align} 
where $\hbar k_\xi=-i \hbar \partial_\xi$ is the momentum operator in $\xi=x,y,z$ direction, $\hbar$ the reduced Planck constant, $m_0$ the bare electron mass, and $\gamma_1=\unit[13.38]{}$, $\gamma_2=\unit[4.24]{}$, and $\gamma_3=\unit[5.69]{}$ the Luttinger parameters for Ge~\cite{terrazosTheoryHolespinQubits2021}. Hamiltonian~\eqref{eq:HLK} also defines the vertical effective mass $m_\perp^{H(L)}=m_0/(\gamma_1\mp 2\gamma_2)$ and in-plane effective mass $m_\parallel^{H(L)}=m_0/(\gamma_1\pm \gamma_2)$.
The spin quantization is given by the growth direction $[001]$ corresponding to out-of-plane $z$-direction. The effect of an external magentic field is included by substituting the momentum with the generalized momentum $\boldsymbol{p}\xrightarrow{} \boldsymbol{p} + e\boldsymbol{A}$, where $\boldsymbol{A}=(2z B_y - y B_z,-2zB_x + x B_z,0)^T/2$ is the electromagnetic vector potential in the Landau gauge~\cite{stanoOrbitalEffectsStrong2019} and $e$ is the electron charge. 

The effect of strain in the Ge well in between the SiGe layers is described by the Bir-Pikus Hamiltonian (see Methods). We assume uniaxial strain ($\epsilon_{xy}=\epsilon_{xz}=\epsilon_{yz}=0$), such that the strain operators become a constant in the different materials. This allows us to describe the effect of strain and an applied electric field in the $z$-direction using the following potential 
 \begin{align}
V_{\perp}(z)=-eF_z \,z - 
\begin{cases}
0, & 0<z<d_i\\
U_{l}, &-d_w<z<0\\
0, &z<-d_w
\end{cases}.
\label{eq:potential}
\end{align}  
Here, $d_w=\unit[18]{nm}$ is the thickness of the strained-Ge quantum well, $d_i=\unit[60]{nm}$ is the thickness of the $\rm Si_{0.2}Ge_{0.8}$ top layer, $F_z$ is the out-of-plane electric field necessary for hole accumulation, and $U_l$ is the band-offset of the heavy-hole ($l=\text{HH}$) and light-hole ($l=\text{LH}$) for the strained Ge layer (see Methods). The SiGe/Ge/SiGe heterostructure is capped by a $\text{SiO}_2$ top interface, modelled as an infinite potential with appropriate boundary conditions $\Psi(z=a_w)=0$. An illustration is shown in Fig.~\ref{fig:Fig2}A.
The in-plane confinement is modelled as a displaced harmonic potential $V_{\parallel}(x,y)= \frac{1}{2} m_{\parallel}^{H(L)} \omega_{0,H(L)}^2(x^2+y^2) + eF_x +eF_y $ with in-plane masses $m_{\parallel}^{H(L)}$ and strength of the harmonic potential $m_{\parallel}^{H(L)} \omega_{0,H(L)}^2 \equiv \frac{\gamma_1+\gamma_2}{m_0} \frac{\hbar^2}{a^4_0}$ with $a_0=\unit[50]{nm}$. In-plane electric fields, $F_x$ and $F_y$, are centred and have average $\braket{F_x}=\braket{F_y}=0$. The magnetic field has a magnitude of $B=\unit[0.1]{T}$ for the simulations presented in this work if not mentioned explicitly, and is applied in the $x$-$z$-plane with an angle $\theta$ between the field direction and x-axis.

The last term in Eq.~\eqref{eq:fullHamiltonian} $H_\text{Zeeman}=2\mu_B \kappa\, \boldsymbol{J}\cdot\boldsymbol{B} + 2\mu_B q (J_x^3 B_x + J_y^3 B_y + J_z^3 B_z)$ describes the interaction between the hole spin and the magnetic field, where $\mu_B=e\hbar/(2m_0)$ is Bohr's magneton, $\boldsymbol{B}=(B_x,B_y,B_z)^T$ the magnetic field, $\boldsymbol{J}=(J_x,J_y,J_z)^T$ the vector consisting of the spin-$\frac{3}{2}$ matrices, and $\kappa=\unit[3.41]{}$ and $q=\unit[0.067]{}$ the isotropic and an-isotropic Zeeman coefficients for Ge~\cite{winklerSpinorbitCouplingEffects2003}.

\subsection{Simulation of g-factor of the ground state}
The total Hamiltonian Eq.~\eqref{eq:HLK} is projected on a set of basis states and then diagonalized numerically. The basis vectors in our simulations consist of product states $\Psi^{H(L)}_{j,k}(x,y,z)=\phi^{H(L)}_j(x,y)\psi^{H(L)}_k(z)$, which are given by independently solving the in-plane and out-of-plane  effective mass  Schr\"{o}dinger equation for the heavy-hole and light-hole bands. The in-plane orbital wave functions are Fock-Darwin states, labelled as $\ket{n,l}$. The z-direction sub-bands of heavy (light) holes $\rm HH_n$ ($\rm LH_n$) have the form of piece-wise Airy functions~\cite{harrisonQuantumWellsWires2016,hosseinkhaniElectromagneticControlValley2020} with Ben-Daniel–Duke boundary conditions (see Methods) $\psi_p(z=a)=\psi_q(z=a)$ and 
$\partial_z\psi_p(z=a)=\partial_z\psi_q(z=a)$ with $(p,q)=(\text{Si}_{0.2}\text{Ge}_{0.8},\text{Ge}),(\text{Ge},\text{Si}_{0.2}\text{Ge}_{0.8})$ and $a=0,-d_w$.
Calculations involving higher orbital states in the realistic heterostructures are computationally expensive.
As the first attempt to simulate sweet spots in the realistic systems, we only considered the effective potentials created in the region of $\text{Si}_{0.2}\text{Ge}_{0.8}$ and $\text{Ge}$, while neglecting the difference of other material parameters such as the Luttinger parameters and Zeeman coefficients.
Fig.~\ref{fig:Fig2} shows the lowest sub-band states in the heterostructure. The wave-functions of the sub-bands can be separated  into states which are localized inside the quantum well, localized at the triangular potential at the surface, or delocalized between well and top-interface. For electric fields $F_z<\unit[3.5]{MV/m}$, there are five heavy-hole states and two light-hole states completely localized inside the quantum well as indicated by the spectrum in Fig.~\ref{fig:Fig2}B and \ref{fig:Fig2}C. We note that with increasing electric fields, first the light-hole states and then the heavy states ``leak" out of the quantum well. The heavy-hole ground state is confined in the quantum well for the electric field lower than $F_z\approx\unit[2.5]{MV/m}$, which marks the upper limit of electric field in this work. We consider three heavy-hole sub-bands and 1 to 57 light-hole sub-bands to simulate the Zeeman splittings of the heavy-hole ground state, which we justify as a sufficient set due to convergence with increasing states. The effective g-factor $g(F_z)$ is then the ratio between Zeeman splitting and the magnetic field strength.

\subsection{Simulation of the dephasing time}
In order to estimate the performance of the planar hole qubits we also compute the effective dephasing times in the presence of charge noise. 
We first model charge noise as random fluctuations of the electric field. For the electric field fluctuations, we assume that the noise follows a $S(f)=A_{\xi}^2/f$ spectral density~\cite{paladinoNoiseImplicationsSolidstate2014, hendrickxFourqubitGermaniumQuantum2021} with $\xi=x,y,z$. To efficiently model the dynamics due to charge noise, we make the following additional assumptions. Firstly, the noise is coupled to the qubit linearly~\cite{ithierDecoherenceSuperconductingQuantum2005,russAsymmetricResonantExchange2015}, secondly, there are no spatial noise correlations, and thirdly, we assume noise in $x$ and $y$ direction to be identical. However, note that these assumptions may break in the presence of alloy disorder, stray strain from metallic gates~\cite{corley-wiciakNanoscaleMapping3D2023}, or extremely close fluctuating charge traps~\cite{wangOptimalOperationPoints2021a}.
Using these assumptions, the pure dephasing time is then given by
 \begin{align}
    T_2^\star(F_\xi) = \frac{\hbar}{ \mu_B \sqrt{ \log(r)} A_\xi \left| \frac{\partial g(F_\xi)}{\partial F_\xi} B \right| }.
\end{align} 
Here, $g(F_\xi)$ is the effective g-factor of the ground state and the bandwidth $r=1.68\,\times 10^9$ is the ratio of the lower and higher frequency cutoff. First-order sweet spots are defined by a vanishing linear noise coupling $\frac{\partial g(F_\xi)}{\partial F_\xi}=0$, thus give rise to exceptionally long dephasing times. 
Because of the finite numbers of basis states included in our simulations and the finite step size in electric field, the g-factor is not completely a smooth function, which gives rise to local variations that overshadow the general trend of $\frac{\partial g(F_z)}{\partial F_z}$. Since these local variations are mostly an artifact of our simulations and our interest lies in the general trend, the interpolated g-factor $g(F_z)$ is fitted to a fourth order polynomial. 

The fluctuation strength of the linear out-of-plane electric field noise is estimated to be $A_z=\unit[3.5]{kV/m}$ inside the quantum well, based on the charge noise estimation~\cite{xueQuantumLogicSpin2022a} from plunger gate fluctuations and Schr\"{o}dinger-Poisson simulation that includes metal/dielectrics gate layers and the germanium heterostructure~\cite{tosatoHighMobilityHoleBilayer2022}, but on the larger side of estimations based on microscopic 3D charge noise simulations~\cite{kepaSimulationChargeNoise2023} in silicon.
Since the g-factor is independent under translation in the $xy$-plane, fluctuating linear in-plane electric fields do not cause any dephasing. However, the hole spin can still be strongly affected by higher-order coupling terms~\cite{boscoSqueezedHoleSpin2021a}.

To provide a realistic comparison, we follow reference~\cite{kepaSimulationChargeNoise2023} and investigate the impact of randomly distributed fluctuating charge traps located at the interface between SiGe and the oxide~\cite{paqueletwuetzReducingChargeNoise2023}. Assuming a continuous metal above the oxide, the potential of a fluctuating charge trap can be well-described by
 \begin{align}
    \delta V_j = \bigg(&\frac{F_c}{|\boldsymbol{r}_j+\delta\boldsymbol{r}_j|}-\frac{F_c}{|\boldsymbol{r}_j|}\nonumber \\
    &-\frac{F_c}{|\boldsymbol{r}_j+\delta\boldsymbol{r}_j+\boldsymbol{r}_\text{m}|}+\frac{F_c}{|\boldsymbol{r}_j+\boldsymbol{r}_\text{m}|}\bigg).
\end{align} 
Here, $\boldsymbol{r}_i=(x_j,y_j,d_i)$ is the location of the charge trap, $\delta\boldsymbol{r}_j$ with $|\delta\boldsymbol{r}_j|=\unit[0.1]{nm}$ is the displacement vector between the two metastable charge states of the fluctuating trap, $\boldsymbol{r}_\text{m}=(0,0,2d_\text{ox})^T$ is the vector pointing to its mirror charge, and $F_c=e/(4\pi\epsilon_0 \epsilon_m)$ is the coupling strength from the Coulomb interaction with $\epsilon_0$ and $\epsilon_m=14.67$ being the vacuum and material permittivity of SiGe. To match a surface charge density of $\unit[1.2\times 10^{-10}]{cm^2}$~\cite{kepaSimulationChargeNoise2023}, we generate 11 randomly positioned fluctuating charge traps in an $\unit[300]{nm}\times\unit[300]{nm}$ area with a random orientation of the displacement vector.
In linear order of coupling strength (see Methods), the total dephasing time is then given in the quasistatic noise limit by~\cite{wangOptimalOperationPoints2021a,shehataModelingSemiconductorSpin2023,kepaSimulationChargeNoise2023}
 \begin{align}
    T_{2,\text{tlf}}^\star=\frac{\sqrt{2}\hbar}{\braket{\sigma_{\delta E}}}
\end{align} 
where $\sigma_{\delta V}$ is the standard deviation of the energy shifts of the individual fluctuators for a given configuration and $\braket{\cdot}$ denotes the average over different of these configurations.
Since the dephasing time as well as the qubit resonance frequency is strongly dependent on the magnitude of the applied magnetic field due to the strong g-factor anisotropy, a comparison of $T_2^\star$ with fixed magnetic field significantly favours small g-factors. To provide a fair comparison of $T_2^\star$ between different magnetic field angles (see Fig.~\ref{fig:Fig5}), we rescale the magnetic field in $T_2^\star$ such that for different magnetic field angles the qubit resonance frequency are equal.


\subsection{Simulation of the Rabi frequency}
Single-qubit gates can be implemented by periodic modulation of gate voltages in proximity of the quantum dot, giving rise to time-dependent electric fields $F_\xi\rightarrow F_\xi + F_{\xi,\text{ac}}\sin(2\pi f_\text{res} t)$  using the cubic Rashba interaction~\cite{terrazosTheoryHolespinQubits2021,sarkarElectricalOperationPlanar2023}. The speed of the operation, the Rabi frequency, can be estimated by (see method~\ref{method:rabi})
 \begin{align}
    \Omega_{\xi,\text{Rabi}}=\frac{1}{h}\left|e F_{\xi,\text{ac}}\bra{0}\hat{\xi}\ket{1}\right|,\label{eq:Rabi}
\end{align} 
where $\hat{\xi}=\hat{x},\hat{y},\hat{z}$ is the position operator and $\ket{0}$ and $\ket{1})$ are the eigenvectors of the qubit states. To provide a fair comparison, we also rescale $\Omega_{\xi,\text{Rabi}}$ such that for different magnetic field angles the qubit resonance frequency are equal.

\begin{figure}[t] 
\centering
    {\includegraphics[width=\columnwidth]{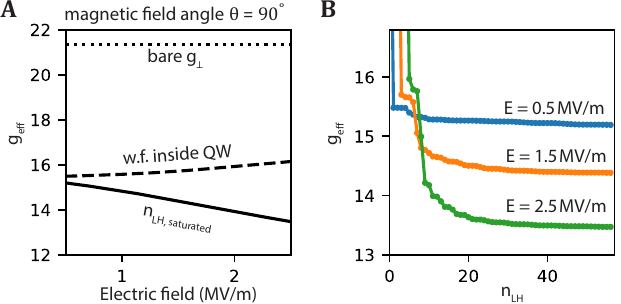}} 
    \caption{\textbf{A}, The out-of-plane g-factor of the ground-state as a function of electric field. The solid curve is the g-factor obtained by including $\rm n_{LH, saturated}=57$ light-hole states in the simulation. The dashed curve is the g-factor obtained by simulating the light-hole states located in the germanium quantum well. \textbf{B}, The g-factor as a function of light-hole level numbers $\rm n_\text{LH}$. Curves in different colors are the results taken at different electric field.  }
\label{fig:Fig3}
\end{figure}


\subsection{Out-of-plane g-factor and convergence behavior}

The out-of-plane g-factor strongly depends on the electric field, as shown in Fig.~\ref{fig:Fig3}A. The g-factor and its derivative changes significantly with the choice of the light-hole states. If we only consider the states in the quantum well, the g-factor is monotonically increasing with respect to the electric field. By incorporating the highly excited light-hole states (up to the $\rm 56^{th}$ excited state in this work), the g-factor changes and is monotonically decreasing with respect to electric field. The zero-derivative point, i.e. the sweet spot, is not observed in the range of electric fields considered here. Applying larger electric fields would result in a ground state that is not located in the quantum well and therefore not considered. Our simulated g-factors match qualitatively with experiments using Hall-Bar measurements at low density~\cite{lodariLightEffectiveHole2019a,lodariLightlyStrainedGermanium2022a}.

We investigate the dependence of the choice of the energy sorted light-hole levels in Fig.~\ref{fig:Fig3}B. The g-factor converges slowly, indicating that the high energy light-hole states are not negligible for the estimation of the g-factor. Large steps in convergence originate from a light-hole state that is localized inside the quantum well, states localized at the top interfaces have minimal impact, and the small steps at larger number originate from delocalized states. We remark that the full 6-band model including the split-off-band (or even more bands) may have to be considered to achieve a higher accuracy of the g-factor.

\subsection{In-plane g-factor}

\begin{figure}[t] 
\centering
    {\includegraphics[width=\columnwidth]{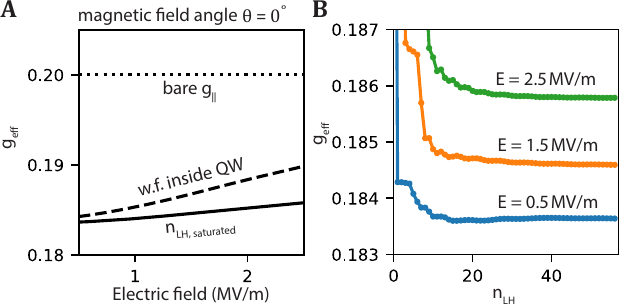}} 
    \caption{\textbf{A}, The in-plane g-factor of the ground-state as a function of electric field. The solid curve is the g-factor obtained by including $\rm n_{LH, saturated}=57$ light-hole states in the simulation. The dashed curve is the g-factor obtained by simulating the light-hole states located in the germanium quantum well. \textbf{B}, The g-factor as a function of light-hole level numbers $\rm n_\text{LH}$. Curves in different colours are the results taken at different electric field.  }
\label{fig:Fig4}
\end{figure}

The in-plane g-factor is plotted in Fig.~\ref{fig:Fig4}A. Compared to the out-of-plane g-factor, the in-plane g-factor is much smaller and it has weaker dependence on the electric field. The g-factor is monotonically increasing with respect to the electric field in both choice of light-hole states, as shown in the dashed and solid curves in Fig.~\ref{fig:Fig4}A. The g-factor dependence of the light-hole levels is plotted in Fig.~\ref{fig:Fig4}B. Our simulation results match the measured g-factors $g=\unit[0.2\pm 0.1]{}$ in devices using the same heterostructure~\cite{hendrickxFourqubitGermaniumQuantum2021}, where the large spread can be attributed to non-circular confinement~\cite{jirovecDynamicsHoleSingletTriplet2022}. The slow convergence is qualitatively similar to the g-factor dependence for out-of-plane magnetic fields. In general, operating planar hole qubits in in-plane magnetic field direction will result in a longer coherence time than operation in out-of-plane magnetic fields.

\begin{figure*}[t] 
\centering
    {\includegraphics[width=\textwidth]{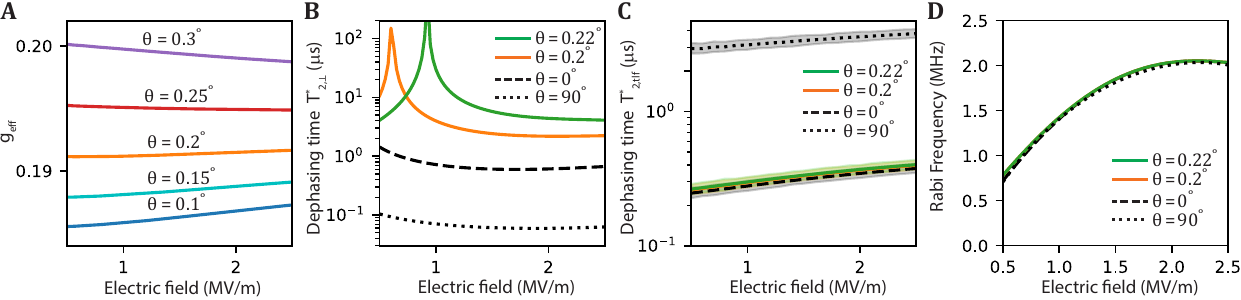}} 
    \caption{\textbf{A}, The g-factor of the ground-state as a function of out-of-plane electric field with different magnetic field angle when considering $\rm n_\text{LH}=57$  light-hole levels. \textbf{B}, Dephasing time $T_{2,\perp}^\star$ due to out-of-plane electric field noise with amplitude $A_z=\unit[3.5]{kV/m}$, plotted as a function of electric field at different magnetic field angle and strength. \textbf{C}, Dephasing time $T_{2,\text{tlf}}^\star$ originating from 11 randomly positioned two-level fluctuators (TLFs) in the quantum dot vicinity, averaged over 200 configurations, and as a function of electric field at different magnetic field angle. The shaded regions represent standard deviations over the simulated configurations estimated through bootstrapping. \textbf{D}, Rabi frequency as a function of electric field at different magnetic field angle and strength. The EDSR driving is at in-plane direction with the magnitude $F_x=\rm \unit[10]{kV/m}$. All four curves are almost overlapping. In plots B-D, the magnetic field strength is chosen such that for each angle the Zeeman splittings are equal (2.5~GHz). 
    }
\label{fig:Fig5}
\end{figure*}

\subsection{Optimal magnetic field angle for out-of-plane electric field noise}
The opposite dependence of the g-factor on electric field for in-plane and out-of-plane magnetic fields, shown in Figs.~\ref{fig:Fig3}A and \ref{fig:Fig4}A, suggests that an optimal field angle exists where the g-factor is first-order insensitive to changes in the out-of-plane electric field. In earlier works, an optimal angle for silicon nanowires was predicted close to $\theta=\arctan(g_\parallel/g_\perp)$~\cite{michalLongitudinalTransverseElectric2021}. Here, we expect the optimal magnetic field angle close to $\theta=\arctan(g_\parallel/g_\perp)/3$ (see Methods). We therefore investigate the angle dependence, shown in Fig.~\ref{fig:Fig5}A. The g-factor as a function of electric field becomes very flat for angles $\theta = \rm 0.2^{\circ} - 0.25^{\circ}$. For certain magnetic field angles, the Zeeman splitting becomes insensitive to electric field fluctuations over a wide range of electric field values, which leads to enhancement of the spin coherence times. Fig.~\ref{fig:Fig5}B shows the estimated dephasing $T_{2,\perp}^\star$ time as a function of electric field, considering fluctuations in $F_z$ at various magnetic field angles. 
From the plot, we find sweet spots at an optimal magnetic field angle of $\theta=\rm 0.22^{\circ}$ if we operate the hole spin qubit at electric fields around $F_z=\unit[1]{MV/m}$. The optimal field angle is decreased if we operate the qubit at lower electric field. We note, that current vector magnets already satisfy the required subdegree precision. In a large-scale germanium quantum processor, each qubit may be brought to its own sweet spot by tuning the electric field to compensate local variations. 

\subsection{Optimal magnetic field angle for fluctuating charge traps}
In Fig.~\ref{fig:Fig5}C we show the dephasing time $T_{2,\text{tlf}}^\star$ originating from randomly positioned two-level fluctuators (TLFs) averaged over 200 configurations and normalized with respect to the Lamor frequency. Our results show $T_{2,\text{tlf}}^\star$ in the range $\unit[200-500]{ns}$ for $|\theta|<\unit[0.25]{^\circ}$, and $T_{2,\text{tlf}}^\star>\unit[1]{\mu s}$ for out-of-plane magnetic fields. 
While a few individual configurations show the emergence of sweet spots in the operation window $\unit[0.5]{MV/m}\leq F_z\leq \unit[2.5]{MV/m}$ that greatly enhance the noise protection (see Fig.~\ref{fig:FigSupp} in Methods), the averaged results do not show such a feature. In contrast to out-of-plane electric fluctuations, for suppressing fluctuating charge traps out-of-plane magnetic field directions are beneficial. We also see an approximately linear relationship between out-of-plane electric field and $T_{2,\text{tlf}}^\star$ for all investigated magnetic fields, indicating a strong impact of higher-order multipole moments. This is in agreement with recent findings that non-separable confinement with respect to in- and out-of-plane can strongly enhance spin-orbit coupling, thus, the susceptibility to charge noise~\cite{martinezHoleSpinManipulation2022}.


\subsection{Total optimal magnetic field angle}
The optimal point of operation is then given by the relative strengths of the different sources of fluctuations and their corresponding dephasing times. For uncorrelated noise the total dephasing time due to charge noise is given by 
 \begin{align}
    \left(\frac{1}{T_{2,\text{tot}}^\star}\right)^2 = \left(\frac{1}{T_{2,\perp}^\star}\right)^2 + \left(\frac{1}{T_{2,\text{tlf}}^\star}\right)^2.
\end{align} 
Since both contributions are of similar order, $T_{2,\perp}^\star \simeq T_{2,\text{tlf}}^\star$, the global optimum depends on the exact configuration of the fluctuating charges, thus be device dependent~\cite{piotSingleHoleSpin2022}. We note, since sweet spots for single charge fluctuators~\cite{wangOptimalOperationPoints2021a,mauroGeometryDephasingSweet2024} or gate electrodes~\cite{piotSingleHoleSpin2022} can be found, a partial sweet spot might be recovered through careful gate calibrations and requires further investigations. Furthermore, for hole qubits in natural Ge quantum wells, dephasing caused by fluctuations of the nuclear spin bath severely limits coherence~\cite{hendrickxSweetspotOperationGermanium2024,mollejoMicrowaveDrivenSinglettriplet2024}

A qubit's quality factor is determined by the number of coherent oscillations within it's decoherence time. Therefore, it is also important to consider how the frequency of coherent oscillations respond to magnetic field angles that yield sweet spots.
Fig.~\ref{fig:Fig5}D shows the Rabi frequency for in-plane driving caused by the cubic Rashba spin-orbit interaction~\cite{terrazosTheoryHolespinQubits2021,boscoSqueezedHoleSpin2021a}. We note that faster Rabi frequencies are accessible using a non-circular in-plane confinement~\cite{boscoSqueezedHoleSpin2021a}, a non-separable confinement~\cite{martinezHoleSpinManipulation2022}, and local strain variations~\cite{abadillo-urielHoleSpinDrivingStrainInduced2022}. Since we do not see a significant drop in Rabi frequency at small angles, the sweet spot allows for fast qubit operations combined with long coherence times. The ability to calibrate each qubit into its own sweet spot with local electric fields can allow compensating local variations through disorder, opening the possibility to a scalable architecture.

\section{Discussion}

In conclusion, we simulated the effective g-factor of hole spins in planar germanium heterostructures and studied its dependence on the electric field, the magnetic field orientation, and the light-hole level numbers. We observed that the excited light-hole levels which are not confined by the quantum well have non-negligible contribution to the g-factor and its derivative with respect to the electric field. 
When including those light-hole levels, we find a tunable sweet spot of the g-factor with respect to out-of-plane electric field if the magnetic field is oriented close to in-plane direction. We note that recent experimental work reporting a sweet spot for holes in silicon FDSOI supports the opportunity for sweet spots for holes in planar germanium~\cite{piotSingleHoleSpin2022}. Decoherence is currently a bottleneck for scaling planar germanium hole qubits~\cite{hendrickxFourqubitGermaniumQuantum2021} thus operating at (scalable) sweet spots may therefore enable the next step in advancing to larger quantum circuits. 

We presented proof-of-principle simulation results by including higher levels and a realistic heterostructure potential. 
Our model can be extended to study the response of hole qubits to decoherence from time-dependent charge noise, g-factor variability from realistic electrostatic and mechanical potentials.

\section{Methods}
\subsection{Derivation of the vertical confinement potential from strain tensor, band offset, and electric field}
The vertical confinement $V_\perp(z)$ of the quantum dot consists of two contributions; alignment of the Fermi-energy of the heterostructure giving rise to a band offset and strain in the quantum well. The band offset is a constant for the different materials and can be experimentally measured or theoretically computed~\cite{schafflerHighmobilitySiGe1997}. Strain is in general a $3\times 3$ strain tensor $\epsilon$ for each band and its effect on the hole states is described by the Bir-Pikusr Hamiltonian. For simplifications, we only consider in this paper the effect of hydrostatic strain and uniaxial strain and ignore all shear-strain components ($\epsilon_{xy}=\epsilon_{xz}=\epsilon_{yz}=0$). Consequently, the Bir-Pikus Hamiltonian becomes diagonal in the heavy-hole and light-hole basis $\ket{j,m_j}=\lbrace\ket{\frac{3}{2},\frac{3}{2}},\ket{\frac{3}{2},-\frac{3}{2}},\ket{\frac{3}{2},\frac{1}{2}},\ket{\frac{3}{2},-\frac{1}{2}}\rbrace$ 
 \begin{align}
    H_\text{PB} = \text{diag}(P_\epsilon+Q_\epsilon,P_\epsilon+Q_\epsilon,P_\epsilon-Q_\epsilon,P_\epsilon-Q_\epsilon)
\end{align} 
with the coefficients
 \begin{align}
    P_\epsilon = -a_V(\epsilon_{xx}+\epsilon_{yy}+\epsilon_{zz}), \\
    Q_\epsilon = -\frac{b_V}{2}(\epsilon_{xx}+\epsilon_{yy}-2\epsilon_{zz}),
\end{align} 
where $a_V$ and $b_V$ are the deformation potentials, which strongly depend on the silicon concentration $x$ in the $\text{Si}_{x}\text{Ge}_{1-x}$ layer of the heterostructure. For $x=20\%$ we use $a_V=\unit[2.0]{eV}$ and $b_V=\unit[-2.16]{eV}$~\cite{terrazosTheoryHolespinQubits2021}.

Since strain is only present in the quantum well and only depends on the band $j=\frac{1}{2},\frac{3}{2}$ and not the sign of the spin, we can rewrite the effect of the band offset and strain as an effective potential of the form
 \begin{align}
V_{\perp}(z)= - 
\begin{cases}
0, & 0<z<d_i\\
U_{l}, &-d_w<z<0\\
0, &z<-d_w
\end{cases},
\end{align}  
where $l=\text{HH},\text{LH}$ denotes the band. Note, that solely due to the uniaxial strain components, the heavy and light-hole degeneracy is lifted inside the quantum well. For our simulations we use the following parameters $U_\text{HH}=\unit[150]{meV}$ and $U_\text{LH}=\unit[100]{meV}$ extracted from~\cite{schafflerHighmobilitySiGe1997} and coincides with the values from~\cite{terrazosTheoryHolespinQubits2021}. By adding a global electric potential $-eF_z z$ originating from the metallic plunger gate on top we end up with expression~\eqref{eq:potential} in the main text.

\subsection{Derivation of the analytical wavefunctions and numerical simulation}
The total Hamiltonian Eq.~\eqref{eq:fullHamiltonian} is projected on a set of basis states and then diagonalized numerically. The basis states for the heavy-hole (light-hole) are product states of in-plane Fock-Darwin wave functions $\phi^{H(L)}_j(x,y)$ and the derived wave-functions in z-direction consisting of piece-wise Airy functions
 \begin{align}
\Psi^{H(L)}_{j,k}(x,y,z)=\phi^{H(L)}_j(x,y)\psi^{H(L)}_k(z),
\end{align} 
with 
\begin{widetext}
 \begin{align}
    \psi^{H(L)}_k(z) = \begin{cases}
c^{H(L)}_{k,1} Ai\left(u_{H(L)}-\epsilon^{H(L)}_k-z/\zeta_0^{H(L)}\right) + c^{H(L)}_{k,2} Bi\left(u_{H(L)}-\epsilon^{H(L)}_k-z/\zeta_0^{H(L)}\right), & 0<z<d_i\\
c^{H(L)}_{k,3} Ai\left(-\epsilon^{H(L)}_k-z/\zeta_0^{H(L)}\right) + c^{H(L)}_{k,4} Bi\left(-\epsilon^{H(L)}_k-z/\zeta_0^{H(L)}\right), &-d_w<z<0\\
c^{H(L)}_{k,5} Ai\left(u_{H(L)}-\epsilon^{H(L)}_k-z/\zeta_0^{H(L)}\right) , &z<-d_w
\end{cases}.
\end{align} 
\end{widetext}
Here, $Ai$ and $Bi$ are the conventional Airy functions, $\zeta_0^{H(L)} =(\hbar^2/(2m_{L(H)}e F_z))^{\frac{1}{3}}$ and $E_\text{tri}^{H(L)}=\hbar^2/(2m_{H(L)}\zeta_0^{H(L)})$ are the effective confinement length and energy of the triangular potential, ${u_{H(L)}=U_{H(L)}/E_\text{tri}^{H(L)}}$ is the effective potential barrier, and $\epsilon^{H(L)}_k=E^{H(L)}_k/E_\text{tri}^{H(L)}$ is the effective eigenenergy of the heavy-hole (light-hole) sub-band $k$. The weighting factors $c^{H(L)}_{k,n}$ are defined via the Ben-Daniel-Duke boundary conditions~\cite{harrisonQuantumWellsWires2016,hosseinkhaniElectromagneticControlValley2020} $\psi_p(z=a)=\psi_q(z=a)$ and  $\frac{1}{m^{H(L)}_{\perp,p}}\partial_z\psi_p(z=a)=\frac{1}{m^{H(L)}_{\perp,q}}\partial_z\psi_q(z=a)$ with $(p,q)=(\text{Si}_{0.2}\text{Ge}_{0.8},\text{Ge}),(\text{Ge},\text{Si}_{0.2}\text{Ge}_{0.8})$ and $a=0,-d_w$. Assuming that the effective masses of the heavy-hole (light-hole) in SiGe are identical to the Ge effective masses, i.e. $m_{\perp,\text{Ge}}^{H(L)}=m_{\perp,\text{SiGe}}^{H(L)}$ and $m_{\parallel,\text{Ge}}^{H(L)}=m_{\parallel,\text{SiGe}}^{H(L)}$, the boundary conditions become independent of the effective mass, and we arrive at the expressions in the main text. We notice that this assumption causes an error of $\unit[5]{\%}$ in $m_{\perp}^H$, $\unit[15]{\%}$ in $m_{\perp}^L$ and $m_{\parallel}^H$, and $\unit[11]{\%}$ in $m_{\parallel}^L$ outside the quantum well. We find the eigenenergies $E^{H(L)}_k$ of the heavy-hole (light-hole) band via the boundary conditions in Eq.~\eqref{eq:potential} following Ref.~\cite{hosseinkhaniElectromagneticControlValley2020} but translate it to a computational task of finding roots of a fifth-order polynomial of the Airy functions. The roots are solved numerically using the \textit{Reduce} function in Mathematica. Afterwards, we check and add missing roots using a bisection algorithm. 

The in-plane orbital wave-fucntions are the solution of a 2D harmonic confinement in the presence of a magnetic field. The general solutions are the Fock-Darwin states
 \begin{align}
    \phi^{H(L)}_{j=(n,l)}(x,y) =& 
     \sqrt{\frac{1}{\pi l^2 }\frac{n!}{(n+|l|)!}} \exp\left(\frac{x^2+y^2}{2a^2_{B,H(L)}}\right) \nonumber\\
    &\times \left(\frac{x^2+y^2}{a^2_{B,H(L)}}\right)^{\frac{|l|}{2}}\mathcal{L}_n^{|l|}\left(\frac{x^2+y^2}{a^2_{B,H(L)}}\right) \nonumber\\
    &\times \exp(-i\, l \arctan(y/x)),
\end{align} 
where $\mathcal{L}_n^{|l|}(\xi)$ are the generalized Laguerre polynomials, $a_{B,H}=\unit[50]{nm}$ and $a_{B,L}=\unit[42.6]{nm}$ are the Bohr radii, and $j$ labels the eigenenergies in ascending order.

For both heavy-hole and light-hole, we use a fixed number of 78 in-plane orbital wave functions. The expression and the integrals between the in-plane orbits are computed analytically. In z-direction, we consider $n_\text{HH}$ heavy-hole sub-bands and $n_\text{LH}$ light-hole sub-bands. We observe that the g-factors changes with $n_\text{LH}$ and saturates as $n_\text{LH}$ increases. The largest $n_\text{LH}$ we consider is 57. Contrarily, the number of heavy-hole sub-bands has a significant smaller impact on the g-factor. The largest $n_\text{HH}$ we consider is 4. The numbers of basis states are $78\times n_\text{HH}$ and $78\times n_\text{LH}$ for heavy-hole and light-hole. The total dimension of the projected Hamiltonian is then given by $n_\text{tot}=156\times (n_\text{HH}+n_\text{LH})$.

We consequently compute the effective g-factor, the ratio of Zeeman splitting to the magnetic field strength, of the heavy-hole ground state by diagonalizing the projected Hamiltonian
 \begin{align}
    g=(E_1-E_0)/(\mu_B B),
\end{align}  
where $E_i$ are the energy-sorted eigenvalues.

To find the electric field dependence of the g-factor, the above procedure is repeated for values of electric field in the interval $F_z=\unit[0.5-3.5]{MV/m}$ with a step size of $\Delta F_z=\unit[5\times 10^{-3}]{MV/m}$. For each electric field value we compute the z-direction sub-bands of the heavy-hole and light-hole, construct the basis states, compute the projected total Hamiltonian Eq.~\eqref{eq:fullHamiltonian}, diagonalize the matrix, obtain the eigenvalues and eigenstates, and finally compute the effective g-factor from the eigenvalues.

To keep the simulation tractable, we truncate the Hilbert space and limit the number of basis wavefunctions $\psi^{H(L)}$. However, due to the dense energy structure of the heavy and light-hole bands with multiple anti-crossings at higher energies (Fig.~\ref{fig:Fig2}), our choice of truncations might miss the respective partner eigenstate at an energy anti-crossing. Together with a finite step size and numerical precision, this leads to small and local fluctuations in the resulting g-factor. While these simulations are not visible in the plots of the g-factors, these fluctuations can affect the derivative $dg(F_z)/dF_z$ and consequently the dephasing time. To avoid these artifacts in our results, we fit the resulting g-factor $g(F_z)$ to a Polynomial in $F_z$ up to fourth order before taking the derivative. We note, that the results are well-approximated by the fitting.

\subsection{Simulation of Rabi frequency}
\label{method:rabi}
Single qubit operations for hole qubits can be implemented by applying an oscillating electric field, $F_\xi \rightarrow F_\xi + F_{\xi,\text{ac}}\sin(2\pi f_\text{res} t)$ with $\xi=x,y,z$, matching the resonance frequency of the qubit $f_\text{res}=2\mu_B g(F_x,F_y,F_z) B/(2\pi \hbar)$. The dynamics of the driven system can be best estimated in the adiabatic frame of Hamiltonian~\eqref{eq:fullHamiltonian}~\cite{messiahQuantumMechanics1961}
 \begin{align}
    H_\text{adiabatic}&=U^\dagger H U -i\hbar U^\dagger \frac{d U}{dt}\\
    &=H_\text{diag} - 2\pi i\hbar f_\text{res} e F_{\xi,\text{ac}}\sin(2\pi f_\text{res} t)  U^\dagger \frac{d U}{dF_\xi},
\end{align} 
where $U^\dagger H U\equiv H_\text{diag}$ contains only diagonal entries.
From the first to the second line, we used $\frac{d U}{dt}=\frac{dF_\xi}{dt} \frac{d U}{dF_\xi}$ with $\frac{dF_\xi}{dt}=2\pi f_\text{res} F_{\xi,\text{ac}}\sin(2\pi f_\text{res} t) $ assuming a linear response and ignoring higher-order terms. The resonant transition amplitude between the qubit states $\ket{0}$ and $\ket{1}$ is then given in the rotating frame by
 \begin{align}
    \bra{0}H_\text{adiabatic}\ket{1}=\pi f_\text{res} e F_{\xi,\text{ac}}(1+e^{4\pi i f_\text{res} t})\bra{0}U^\dagger\frac{\partial U}{\partial F_\xi}\ket{1}.
\end{align} 
By ignoring the counter-rotating term, the so-called rotating wave-approximation, we end up with expression~\eqref{eq:Rabi} of the main text. Conveniently, this method requires only knowledge about the instantaneous eigenvectors of the qubit space. The Rabi frequency is then given by 
\begin{align}
    \Omega_{\xi,\text{Rabi}}= \frac{2}{h} |\bra{0}H_\text{adiabatic}\ket{1}|.
\end{align}
If we further use the linearity of the driving , i.e., $H_\text{tot}= H + e F_{\xi,\text{ac}}\sin(2\pi f_\text{res} t) x$, the upper expression can be recast into the more familar expression
 \begin{align}
    \Omega_{\xi,\text{Rabi}}= \frac{1}{h} |e F_{\xi,\text{ac}} \bra{0}\hat{\xi}\ket{1}|,
\end{align} 
where $\hat{\xi}=\hat{x},\hat{y},\hat{z}$ is the corresponding position operator.

\subsection{Optimal magnetic field angle for out-of-plane fluctuations}
The emergence of an optimal magnetic field angle can be derived from Hamiltonian~\eqref{eq:fullHamiltonian} of the main text. While this derivation can be easily generalized to arbitrary magnetic fields, we pursue a magnetic field in the $xz$-plane $\boldsymbol{B}=(B\cos(\theta),0,B\sin(\theta))^T$. To diagonalize the heavy-hole state sector we apply the unitary rotation $U=e^{-i \phi \sigma_y/2}$ with $\sigma_z$ being the Pauli matrix acting only on the heavy-hole space and 
 \begin{align}
    \phi = \arctan\left(\frac{4\kappa+9q}{2q}\tan(\theta)\right) = \arctan\left(\frac{g_\perp}{g_\parallel}\tan(\theta)\right).
    \label{eq:method_phi}
\end{align} 
Here, $\kappa$ and $q$ are the isotropic and an-isotropic Zeeman coefficients and $g_\perp=6\kappa+27q/2$ and $g_\parallel=3q$ are the out-of-plane and in-plane pure heavy-hole g-factors. While the angle $\theta$ describes the rotation of the magnetic field, the angle $\phi$ describes the rotation of the heavy-hole quantization axis. Minimal variation of the g-factor is then expected to be close to $\phi=\unit[45]{^\circ}$ where the orbital contributions from in-plane and out-of-plane magnetic fields compensate each other~\cite{michalLongitudinalTransverseElectric2021}. From our simulations, we can see that the ratio of the slopes $\frac{\partial g(F_z)}{\partial F_z}$ normalized to equal qubit frequencies for $\theta=\unit[90]{^\circ}$ and $\theta=\unit[0]{^\circ}$ are not equal, therefore we end up with $\theta_\text{opt}\approx \arctan(g_\parallel/g_\perp)/3$.
\begin{figure*}[t] 
\centering
    {\includegraphics[width=\textwidth]{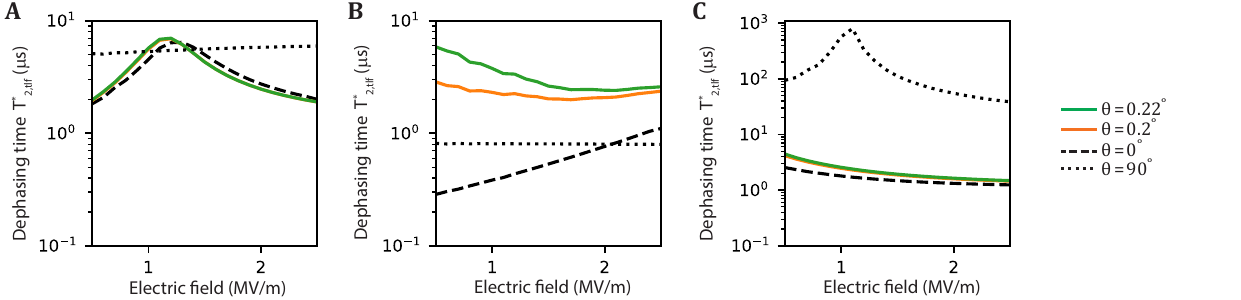}} 
    \caption{Dephasing time $T_{2,\text{tlf}}^\star$, caused by three different TLF configurations, as a function of out-of-plane electric field with different magnetic field angles. \textbf{A}. Emergence of a sweet spot for magnetic fields with small $\theta$. The sweet spot is robust against small changes in magnetic field orientation. \textbf{B}. Emergence of incomplete sweet spot features that are highly sensitive to magnetic field orientations. \textbf{C}. Emergence for a sweet spot for out-of-plane magnetic field similar to the one reported in Ref.~\cite{wangOptimalOperationPoints2021a}.}            
\label{fig:FigSupp}
\end{figure*}

\subsection{Optimal magnetic field angle for fluctuating charge traps}
The potential caused by a single charge trap, approximated as point-charge, is given by the Coulomb potential. The potential difference of a two-level fluctuator (TLF) subject to screening from the metal gates (here assumed to be continuous) is given by
 \begin{align}
    \delta V_j = \bigg(&\frac{F_c}{|\boldsymbol{r}_j+\delta\boldsymbol{r}_j|}-\frac{F_c}{|\boldsymbol{r}_j|}\nonumber \\
    &-\frac{F_c}{|\boldsymbol{r}_j+\delta\boldsymbol{r}_j+\boldsymbol{r}_\text{m}|}+\frac{F_c}{|\boldsymbol{r}_j+\boldsymbol{r}_\text{m}|}\bigg).
\end{align} 
The first two terms are the potentials caused by the two meta-stable states of the two-level fluctuator with the remaining terms being their image charges. Here, $\boldsymbol{r}_i=(x_j,y_j,d_i)$ is the location of the charge trap, $\delta\boldsymbol{r}_j$ with $|\delta\boldsymbol{r}_j|=\unit[0.1]{nm}$ is the displacement vector between the two metastable charge states of the fluctuating trap, $\boldsymbol{r}_\text{m}=(0,0,2d_\text{ox})^T$ is the vector pointing to its mirror charge, and $F_c=e/(4\pi\epsilon_0 \epsilon_m)$ is the coupling strength from the Coulomb interaction with $\epsilon$ and $\epsilon_m=14.67$ being the vacuum and material permittivity of SiGe. 

We consider 11 randomly positioned charge traps that serve as two-level fluctuators (TLFs) in a $\unit[300]{nm}\times \unit[300]{nm}$ area drawn from a uniform distribution. We furthermore draw the vector connecting the two meta-stable states of the fluctuator $\delta\boldsymbol{r}_j$ from a uniform 3D vector with fixed length $|\delta\boldsymbol{r}_j|=\unit[0.1]{nm}$. The corresponding potential for a given configuration reads
 \begin{align}
   V_{\boldsymbol{b}} = \sum_{j=1}^{11} \delta V_j \bigg|_{\delta\boldsymbol{r}_j\rightarrow b_j \delta\boldsymbol{r}_j }.
\end{align} 
Here $\boldsymbol{b}$ is a binary vector indicating the current state of each TLF, i.e. $0$ for not displaced and $1$ for displaced. For example, $(0,\cdots, 0)^T$ represents all charge traps in their original position. 
To get the average fluctuations, we compute for each state of the TLFs the corresponding qubit energy shift
 \begin{align}
   \delta E_{\boldsymbol{b}_k} = \bra{0}V_{\boldsymbol{b}_k}\ket{0}-\bra{1}V_{\boldsymbol{b}_k}\ket{1},
\end{align} 
where $\ket{0}$ and $\ket{1}$ are the qubit states. To speed up the computation, we use instead a series expansion of the upper expression up to $6^{\rm th}$-order in $x$ and $y$ and up to second order in $z$. In our simulations, we make use of our analytical expressions and compute the matrix elements from a general polynomial and substitute later the actual values.

The total fluctuations caused by the TLFs are consequently given by the root-mean-square with respect to the TLF states
 \begin{align}
   \sigma^2_{\delta E} = \frac{1}{N^2}\sum_{k}\delta E_{\boldsymbol{b}_k}^2,
\end{align} 
where $N$ is the number of TLF states. In our simulations, we linearize the problem and neglect TLF states with more than one excitation. This is a good approximation~\cite{shehataModelingSemiconductorSpin2023,kepaSimulationChargeNoise2023} and becomes exact if $\delta E_{(\cdots,1,\cdots,1\cdots)}=\delta E_{(\cdots,1,\cdots,0\cdots)}+\delta E_{(\cdots,0,\cdots,1\cdots)}$ and if there is no correlation between the TLFs.

As a final step, we repeat the upper steps for 200 configurations of the 11 TLFs and average over them.

\subsection{Optimal magnetic field angle for selected individual fluctuating charge traps}
Fig.~\ref{fig:FigSupp} shows the dephasing time caused by a few selected TLF configurations as a function of out-of-plane electric field for different magnetic field configurations. Depending on the configuration, sweet spots can appear for small $\theta$ angles (Fig.~\ref{fig:FigSupp}A), just out-side the window of investigation (Fig.~\ref{fig:FigSupp}B), and also for $\theta=\unit[90]{^\circ}$  (Fig.~\ref{fig:FigSupp}C). 

\paragraph*{Acknowledgements}
We acknowledge helpful discussions with D. DiVincenzo, M. Friesen, N. Hendrickx, S. Philips, H. Tidjani, A. Tosato,  L. Vandersypen, and all members of the Vanderypen, Veldhorst, and Scappucci group. C.W. M. V. acknowledge support by the European Union through ERC Starting Grant QUIST (850641). Research was sponsored by the Army Research Office (ARO) and was accomplished under Grant No. W911NF- 17-1-0274 and W911NF-23-1-0110. The views and conclusions contained in this document are those of the authors and should not be interpreted as representing the official policies, either expressed or implied, of the Army Research Office (ARO), or the U.S. Government. The U.S. Government is authorized to reproduce and distribute reprints for Government purposes
notwithstanding any copyright notation herein. M.R. acknowledges support from the Netherlands Organization of Scientific Research (NWO) under Veni grant VI.Veni.212.223.

\paragraph*{Author contributions}
M.R.-R. and M.V. conceived and supervised the project. M.R.-R. developed the theoretical model and C.-A.W. performed the simulations with input from E.E. and M.G..  M.R.-R. and C.-A.W. performed the analysis with input from G.S. and M.V.. C.-A.W wrote the manuscript with input from G.S., E.E., M.G., M.V., and M.R.-R..
\paragraph*{Competing interests}
The authors declare no competing interests.
\paragraph*{Data availability}
Simulation software and data analysis scripts supporting this work are available at \doi{10.5281/zenodo.6949625}.
\bibliography{literature_database}

\end{document}